\documentclass{article}
\pdfoutput=1
\usepackage{geometry}               
\usepackage[parfill]{parskip}    
\usepackage{graphicx}
\usepackage{amssymb}
\usepackage{epstopdf}
\usepackage{verbatim}
\usepackage[toc,page]{appendix}
\usepackage{courier}
\newenvironment{mylisting}
	{\begin{list}{}{\setlength{\leftmargin}{1em}}\item\scriptsize\bfseries}
	{\end{list}}

\title{Neurally Implementable Semantic Networks}
\author{Garrett N. Evans\footnote{gne101 @ psu.edu}\
        \ and John C. Collins\footnote{collins @ phys.psu.edu} \\
        \\ \emph{104 Davey Lab, Penn State University, University Park, PA 16802, USA} }

\date{March 18, 2013}

\begin{document}
\maketitle

\begin{abstract}

We propose general principles for semantic networks allowing them
to be implemented as dynamical neural networks. 
Major features of our scheme include: (a) the interpretation that each 
node in a network stands for a bound integration of the meanings of all 
nodes and external events the node links with; (b) the systematic use of 
nodes that stand for categories or types, with separate nodes for 
instances of these types; (c) an implementation of relationships
that does not use intrinsically typed links between nodes.

\end{abstract}

\section{Introduction}

Semantic networks are an important and successful means for
representing knowledge in computer systems and other types of
information technology. Endeavors in which they have proven beneficial include: (a)
natural language processing and image analysis \cite{Wordnet2010,
  reERNEST}; (b) commerce, in which they aid communication
among experts with different specialties and show promise for database and web
search, organization and mapping \cite{EnterpriseOntology,
  McGuinness}; and (c) diagnosis and identification of
illness outbreaks \cite{reSAPPHIRE}. The nodes of a semantic network
represent ideas, and the 
structure of links between nodes allows the network to represent meaning that
involves these ideas.

It is natural, given a semantic network, to try to implement it
dynamically in a neural network.  One motivation is that semantic
networks can be used to code human knowledge, for which the biological
implementation is neural. It is conceivable then that the brain may be structured
in such a way that readily distinguishable patterns
of neural activity correspond to nodes in a semantic network and that synapses
or collections of synapses between neurons correspond to links.

However, a fundamental difficulty in finding an adequate neural
implementation of a semantic network is that neurons and synapses have
no intrinsic semantic content. In a semantic network, the nodes have meaning, and
the links typically have a variety of semantic types to code different types of relationship;
the semantics are often specified by linguistic labels.
But neural firing is just an electrical signal, and
synapses are essentially untyped links between neurons.
 
The purpose of the present paper is therefore to show how to formulate
semantic networks that do not intrinsically rely on labels or semantically typed links
and that are suitable for neural implementation. 
We propose a set of criteria to achieve this and then show how they may be applied along with the consequences that they entail.  The criteria are:
\begin{enumerate}
\item \label{criterion:nodeMeaning} The meaning of nodes should come
  from links they share with other nodes or with environmental events,
  not from labels. The ultimate grounding of the semantics of the
  nodes then arises, if indirectly, from neurons/nodes that
  have direct interaction with the external world. Any node labels used in diagrams should only
  reflect meaning that is acquired in this way.\footnote{One use of
    labels on nodes arises when one examines a subnetwork.  Then the
    grounding of some nodes is external to the subnetwork, and labels
    can be used to give the results of the external grounding.
    Nevertheless, as a general principle, it is important that the
    final grounding for node semantics in a complete neurally
    implementable semantic network should lie in connections to the
    environment.}

\item \label{criterion:diffNodes} Different ideas recognized by a network need separate nodes, 
even when the ideas are related. By \emph{idea}, we mean any possible subject of thought.

\item \label{criterion:linkSemantics}  Links in a network may have direction and weight but no special semantics attached.

\item \label{criterion:expressRelations} Nonetheless, relational structure must be expressed by the network.

\item \label{criterion:DRY} Excessively duplicative encoding should be avoided.

\end{enumerate}

One of the most central consequences of these criteria concerns generic concepts or \emph{types} and specific \emph{instances} of types. Each instance of a type, e.g., of a flower, is a distinct idea, so our second criterion requires that there be a node for it. The fact that these instances all hold something in common is expressed by a link between each \emph{instance node} and a \emph{type node} for the generic concept itself, which represents what is common among the instances.

\begin{figure}
\centering
\includegraphics[scale=0.5]{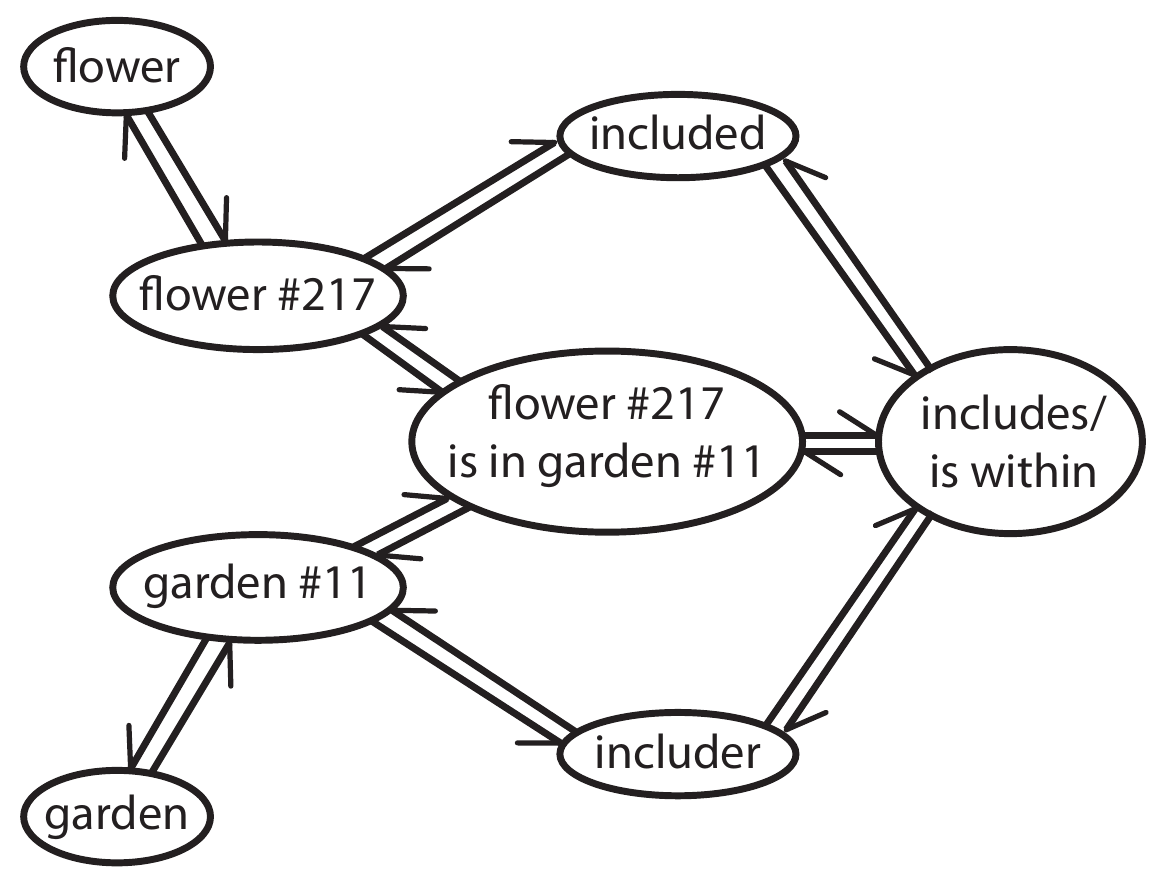}
\caption{
A small subnetwork demonstrating type and instance nodes including relationship type and instance nodes. ``Flower \#217 is in garden \#11'' is an instance of the `includes/is within' relation type. The `flower \#217' node is an instance of the `flower' entity type.
}
\label{fig:flowerGarden}
\end{figure}

In general, we will consider the meaning of any node in a semantic network
to be an \emph{integration} of the
meanings of nodes it shares links with along with any environmental
events it links with as well. We will represent a relationship between two
nodes, not by a simple link, but by a pair of links
with an intermediate node standing for the relationship itself, as well as nodes standing for 
different positions in the relationship when needed.  The intermediate node usually codes a
particular instance of a particular kind of relationship. To express this, as with any
instance node, the relationship instance node
needs a link to a type node for the relationship
type it belongs to.  The simple subnetwork in Fig.~\ref{fig:flowerGarden}
illustrates what we mean.

Our systematic use of instance and type nodes for concepts and
relationships also provides a way of meeting our fifth criterion, 
which corresponds to what in computer
software engineering is called the DRY principle, ``Don't Repeat
Yourself.''  In semantic network terms, this implies that the node and
link structure by which a generic concept is defined should normally
occur just once within the network: at type nodes.  The links
between instance nodes and type nodes allow the generic meaning to be inherited
by instance nodes. Without this use of type nodes, not only will there
be considerable inefficiency, but corrections to the meaning of the
concept can only be learned slowly and inflexibly, if at all.
Duplication should only be used for strong reasons, with special
mechanisms to keep semantics synchronized between duplicates under the
influence of future learning.

The single type node for each concept has further links with the
correct nodes or external events necessary to code the concept's
meaning.  This semantics is transmitted to instance nodes by the links
between the type node and the instance nodes.  Applying this
principle to relationship and relationship position nodes will mean
using type nodes for relationship types, e.g, `includes/is within', which
will link with specific instances of the types, such as ``The flower
is in the garden.''  A similar structure will be used for positions in
relationships, for instance, `includer' and `included' in the case of the present
example.  We will explain these principles in greater depth in Sec.
\ref{Sec:Framework}.

An important consideration regarding our method arises due to our taking it for granted
that learning occurs over the lifetime of any naturally occurring
neural system implementing a semantic network as the system
continually encounters new \emph{entities}, that is, concrete stimuli.
The system will benefit from organizing these into types, and it therefore has a
lifetime need to create new types, including types for newly learned
relationships.  Learning also includes additions to both factual and
episodic (i.e., event) memory.  Presumably in a practical system, it
can further be useful to remove nodes and links that get little usage.
While neurons themselves do not get eliminated or created very often
in the brain, synapses among neurons are plastic and can be rewired so
that the neurons stop implementing nodes and links that are not useful and
start implementing nodes and links that are. While our model will use a
considerable proliferation of
nodes to express knowledge compared with many semantic networks seen
in the literature, the increased number of nodes may be the price
to pay for a neurally implementable system that is to deal with a
human level of knowledge and which is sufficiently flexible in its
learning ability. Also, as we say, nodes need only be as permanent for
a system utilizing this methodology as is the knowledge they help represent.

Our endeavor here is motivated by and draws upon an array of previous experimental and theoretical work including cell assembly theory due to Wickelgren and others \cite{Wickelgren92, Wickelgren99}; semantic network theory, which is reviewed in \cite{Sowa} and in \cite{Luger228}; and cognitive architecture methods due to D\"orner, Bach and colleagues \cite{ Bach, BachEtAl, Bartl&Dorner}. Observations by Quiroga and others  \cite{HahnloserKozhevnikovFee, QuirogaEtAl2005, QuirogaEtAl2008, QuirogaEtAl2009} of cells responding robustly and specifically to conceptually oriented stimuli also serve to indicate that a theory ascribing an explicit conceptual role to neurons or sets of neurons may be accurate. Our intention is to build in the direction of a comprehensive theoretical framework incorporating these related results. We will discuss their significance along with some other important issues in the following section.

As a final matter of introduction, we would like to refer the reader to Appendix \ref{App:Glossary} for a selected glossary of some of the more important terms as defined and used in this paper.

\section{Background}

\subsection{Cell Assemblies}

Cell assembly theory was initiated by Hebb, who suggested there might be overlapping groups or ``assemblies'' of interconnected neurons which ``underly'' the perception of various kinds of objects such as shapes \cite{Hebb49}. Hebb's assemblies are interconnected, by way of synapses, with each other and with neurons representing the constituent features of the perceptions, thus forming a neurally implemented network of perceptual ideas.

Since Hebb, many authors including Wickelgren \cite{Wickelgren92, Wickelgren99}, Palm \cite{Palm81} and Leg\'endy \cite{Legendy67} have expanded cell assembly theory. These authors have presented suggestions for how neural cell assemblies might be structured and how they could adaptively come to represent new and useful percepts. Many of them, particularly Wickelgren, also adopt an extension to the scope of Hebb's proposal: that neural assemblies exist not just for perceptions but for all mental ``ideas'', for example, ``the word `dog', the image of a particular dog, the concept of a particular individual dog, the concept of a member of the set of dogs, the concept of the set of all dogs, the proposition that dogs eat meat, etc.''~\cite{Wickelgren92}. These neurally implemented networks of \emph{ideas}, a term we have borrowed from Wickelgren, qualify as semantic networks because, for all of these authors, assemblies respond to the stimuli they do and thus have the meaning they do because of synaptically implemented interconnections between assemblies: the network topology establishes meaning.

To the best of our knowledge, cell assembly theorists do not have a complete proposal for how relational structure within and among ideas gets represented by networks of cell assemblies. For example, given the ideas, `father', `mother' and `child', links connecting these ideas may show that they are related to each other, but the subtleties, particularities and differences involved in these relationships are unclear: Mother is related to father in a different way than mother is related to child and likewise for father and child, but simply linking the ideas together does not express this. Hebb suggests that assemblies standing for sensory percepts such as shapes could represent the structure of the percepts by way of links to motor programs that move sensory organs in such a way as causes the percept to be recognized only if its components are situated in the correct way \cite{Hebb49Ch5}. It is unclear, though, how motor programs would enable the representation of relational structure among more abstract ideas like mother, father and child.

\subsection{Spreading activation models}

\begin{figure}
\centering
\includegraphics{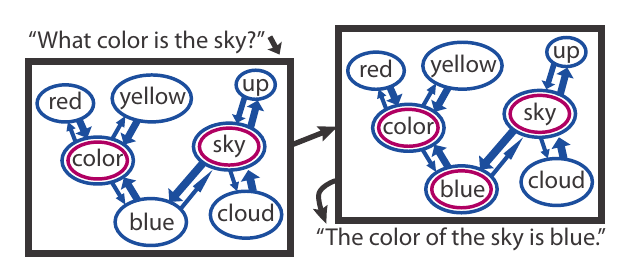}
\caption{Simple spreading activity network. Magenta indicates node activity.}
\label{fig:spread}
\end{figure}

Models of cognition as spreading activation on semantic networks have a kinship with cell assembly theory and are familiar \cite{CollinsLoftus,Luger228,Quillian67}. Fig.~\ref{fig:spread} shows a simple example of this type of model. In the figure, the network is primed by the question, ``What color is the sky?'', which initiates activity in the `color' node and the `sky' node. Most of the remaining nodes in the network receive connections from either the `color' node or the `sky' node but not both, which we take to be insufficient for transmission of activity. But because the `blue' node receives a dynamical connection from both active nodes, the activity spreads to the `blue' node as shown in the second figure. The combined activity of the `color', `blue', and `sky' nodes is taken to signify the recollection, ``The color of the sky is blue,'' which provides the answer to the query. It is assumed that this internal representation can then be translated to an output mode.

The dynamics of Fig.~\ref{fig:spread} seems naturally suited for translation into the flow of action potentials within a neural network, providing a basis for a neural theory of information storage and recovery. The essential structures are the same: units which can be active or inactive and which can transmit activity to other units that they are connected to. The primary task for such a reductive effort would be to determine the neural structures which nodes should correspond to. As with cell assembly theory, however, matters become more difficult when one wishes to represent \emph{relationships} existing among the ideas represented by nodes, e.g., that clouds are \emph{in} the sky and that color is a \emph{type} that blue \emph{belongs} to. As we have said, representing relationships with semantic attachments to links, while common in semantic networks, is problematic for neural implementability.

\subsection{Neurally Implementable Expression of Relationships}

Since neural networks are made up of dynamical connections among neurons, with spiking activity in pre-synaptic neurons (usually) influencing spiking activity in post-synaptic neurons, if a semantic network schema is to be implementable in a simple and straightforward way by neural networks, it makes sense for the schema to utilize connections to which this same basic interpretation can be applied. This is not the case, though, for semantic networks that use links with specialized semantics, as is popular. For example, Quillian's seminal work \cite{Quillian67} uses five different types of link denoting, respectively, subsumption (the `is-a' relation), modification, disjunction, conjunction, and subject-object-relation triads. If one were to attempt to translate a strategy similar to this into neural terms, it would be difficult, for example, to distinguish how subsumption-related influence of activity should be different from modification influence of activity, etc. This is why we restrict our model to links which fit the standard mathematical definition for links in directed networks: links can have direction and weight but no semantic attachments (see criteria on pg. \pageref{criterion:linkSemantics}).

The problem is, how can one represent relational structure among ideas using only such links. Strategies such as John Sowa's Conceptual Graphs \cite{Luger228} \cite{Sowa} accomplish this by placing, between two related nodes, a pathway involving two links and an intermediate node signifying the relationship the nodes share. For example, this would amount to placing an `is-in' intermediate node along the pathway from `cloud' to `sky' in Fig.~\ref{fig:spread} to express that clouds are found in the sky. The approach essentially moves relational specification from links themselves to intermediate nodes, introducing into the network many nodes with closely related semantics, including one `is-in' node for each instance of an entity or type being part of another.

Because relationships between ideas are themselves ideas, they deserve
nodes according to the second of our suggested criteria
(pg.~\pageref{criterion:diffNodes}) for neurally implementable semantic networks,
that distinct ideas recognized by a network have distinct nodes.
Therefore, expressing relationships with intermediate relation-specifying nodes
is a fit for us. With the adoption of this mechanism arises another issue, however.
Recall our first criterion: that the meaning of nodes cannot come from labels applied
to nodes, as in Conceptual Graphs and similar methods, but must instead come from link partners.
The most straightforward way to satisfy this principle in the case of
intermediate relation-specifying nodes is to say that each
of these nodes should have, in addition to its connections with the nodes it relates,
links coupling it with a set of
nodes which define the relationship itself.  This will involve a tremendous
amount of information duplication, though.  Encoding, via linkages, the full
semantics of the `is-in' relationship, which is highly abstract, separately for each node
in a network signifying that one idea is included in another would grossly
violate our fifth criterion (pg.~\pageref{criterion:DRY}) for our method, which is the avoidance
when possible of repetitiveness.

Our solution is to assign repeated relational meaning to intermediate
relation-specifying nodes, not on an individual basis, but by
expressing that such nodes are each an instance of a single relationship type,
the `is-in' relationship type in the case of the example we have been using.
Since relationship types, such as `is-in', are ideas we
expect networks to have the capacity to recognize, nodes standing for these
relationship types should exist according to our second criterion.
A node standing for the relationship type will encode what it means for any given
situation to be an `is-in' or any other kind of
relationship by way of links to nodes providing
this definition. This allows the intermediate nodes existing between ideas
which possess the relationship to encode that they
belong to the type and thus inherit its associated relational meaning by way of a
single link shared with the relationship type node. 
Because the intermediate
relation-specifying nodes serve as instances of the relationship type, we will
refer to them from now on as \emph{relationship instance nodes}.

Joscha Bach and Dietrich D\"orner use a similar technique to this in their Psi and MicroPsi cognitive architectures \cite{Bartl&Dorner,BachEtAl,Bach}. In the Psi model, gating nodes are used as intermediaries between nodes which are related in one of a fixed set of four ways, viz., the ``is-in'' relationship, the ``succession'' sequential relation and the inverses of these. Gating nodes establish their role via a link to a node standing for the relation type they instantiate. Thus they have exactly the same structure as relationship instance nodes in our framework.  Bach showed that this type of dynamic node architecture can autonomously learn and produce effective behavior \cite{Bach}. The important difference between Psi and our approach is that we do not limit the set of relations that can be given a relationship type or instance node. We expect that, in the same way that an organism learns suitable types for the entities common and relevant to its environment by experience, it will acquire types for the \emph{relations} common and relevant to that environment as well, which will be numerous. Once established, relationship type nodes can be put directly to use expressing relations between entities and types through the gating (relationship instance) node method.

\subsection{Experimental Support and the Neural Implementation of Nodes}
\label{ExpSupportNodes}

Two major experimental findings lend support to the possibility of semantic-network-like encoding of information in the brain and have led to our interest in developing this new framework. One is the observation of neurons such as those found in the medial temporal lobe (MTL) which selectively respond to the presence of abstract percepts, including specific noteworthy people. Among these are neurons responding not only to the visual presentation of the individual's face under different lighting conditions and differing spatial orientations but also to presentation of the person's written \cite{QuirogaEtAl2005,QuirogaEtAl2008} and spoken \cite{QuirogaEtAl2009} name and even to thought concentration \cite{CerfEtAl} on the person. Similar cells are found in MTL which selectively respond to various images of a famous landmark.

Neurons with related properties have been found in songbird higher forebrain nucleus HVC. These neurons are observed to fire in conjunction with specific points within vocalization sequences found in the bird's singing repertoire. One type of neuron found in this area (HVC$_{\rm RA}$) becomes active only when the (awake) bird is singing \cite{HahnloserKozhevnikovFee}. Neurons belonging to another class (HVC$_{\rm X}$) activate in correspondence with the same position in a vocal sequence regardless of whether the bird is hearing the song or producing it (even without auditory feedback) \cite{PratherEtAl}.

These two sets of findings demonstrate the existence of neural activity in brains which correlates well with \emph{abstract} concepts, i.e., recognized types for stimuli delineated by something other than base-level features. In the case of the MTL results, the category is stimuli or thought involving a certain famous person or landmark. In the case of the HVC$_{\rm X}$ neurons, it is a position in a sequence, regardless of whether that sequence is being detected or produced. Since `abstract concepts' are a characteristic sort among things commonly appearing as nodes in a semantic network, these findings are consistent with the hypothesis that the brain uses neurons to implement the nodes of a semantic network\footnote{Recorded neurons may or may not, themselves, be incorporated into a node standing for the concept they are observed to respond to. They may, for example, belong to a node standing for a closely related idea; the point is the response indicates that neural representation of the concept is happening \emph{somewhere.}} in an explicit way \cite{Quiroga2012}. Given the data, there are multiple options for the manner in which the nodes are implemented neurally, a few of which we will enumerate.

The simplest possibility is that there are separate groups or assemblies of
neurons implementing single nodes
representing ideas in a semantic network with synaptic
connections between assemblies implementing links between nodes. Such
assemblies then stand for the particular bundle
of elements that is a given idea. The
neurons then give a purely local representation of the ideas. Wickelgren
refers to this as \emph{giant neuron coding} because the neurons
representing an idea act together as a single collective unit---engaging in
sustained firing together or not at all \cite{Wickelgren99}. The neurons detected above
could be members of giant neuron assemblies in this case. 

A next-order complexity implementation is for nodes to correspond to distinct but overlapping sets (still considered assemblies) of neurons. Wickelgren prefers this implementation, which he calls \emph{overlapping set coding}, because it greatly increases the number of nodes that can be implemented by a network of neurons \cite{Wickelgren99,Legendy67,Palm81}. Non-independence among synaptically implemented links could be an issue for this type of coding, which sparse coding may remedy \cite{Foldiak,RollsTreves}.
Taking a step further in complexity, it is also a
possibility that nodes could correspond to specific recurring
spatio-temporal patterns of activity among neurons, cf.~\cite{Izhikevich2006}.

In these last two cases, the
neurons observed to respond to a single unique concept in the experiments described above would also robustly respond to a few other unrelated concepts that were not presented. Both of the higher order complexity hypotheses, in addition to the simpler one, would predict the detection of neurons with the properties of those found in the studies we mentioned. Our concentration here, however, is not on explicating a single specific way in which nodes are implemented neurally although this is certainly an important task. Rather, we are most concerned with providing a semantic network methodology that lends itself to implementation in terms of neurons in one or more possible ways. We will leave the specifics to future work.

\subsection{Synchrony and the binding problem}
\label{subsection:Synch_Binding}

The second key result inspiring our efforts is that neurons have
been observed to exhibit time-synchronized firing in the event that their
receptive fields correspond to the component features of a stimulus
that is being perceived as a single \emph{entity} \cite{Singer1999}, that is, as
a concrete conglomerate of one or more features or sub-entities. Because incorporation into an
integrated composite can be seen as a kind of semantic relation among individuals
(which, as we will see, forms the basis for \emph{all} semantic relation in our proposal), 
these results demonstrate an experimentally observed connection between synchronous firing
among neurons and semantic relatedness among the entities which the
neurons code for. Time-synchronous firing
among neural populations is known to be causally related to synaptic
connectedness \cite{Sporns} and may provide a way for synaptically encoded semantic
relatedness to be retrieved for processing (a synchrony code).
Conversely, perceived relational structure within a stimulus, if
encoded by synchrony, could potentially be stored into memory via
Hebbian synaptic plasticity processes producing synaptic connections
that reflect the synchrony.

Encoding the incorporation of multiple features into a singly entity by way of temporally correlated spiking activity is connected to the well-known theoretical problem for
neural representation known as the \emph{binding problem}---see,
e.g., Ref.~\cite{Malsburg95}. Features which are considered components
of a single entity are said to be \emph{bound}.  For example, if a neural
system is to represent a situation involving a red circle and a blue
square, it is not sufficient for the features circle, square, red and
blue to be represented as merely present by activity in the
associated nodes.  Something must be done to indicate that the circle
and red features are bound to one another as are the square and blue
features.  Otherwise, the organism utilizing this neural system will
be unable to distinguish the case of a red circle and blue
square from that of a blue circle and red square.

While a synchrony code can account for the representation of binding with respect to an organism's current situation, binding must also be achieved within and among its stored memories of past situations.  Neural and synaptic devices are necessary for this. Useful in this regard is what we shall refer to as a \emph{binder unit}. A binder unit is an element of memory that serves to integrate a particular set of features or other components into a unified whole. In the present context, we will assume this to be accomplished by way of synaptic connections between the neurons representing the components and the those that constitute the binder unit. The neural implementation for the binder unit becomes activated in the presence of a suitable level of collective activation in the neural implementations for the components (being perhaps especially sensitive to synchronized spiking). This allows the unit to recognize the integration\footnote{The emphasis is on \emph{integration} rather than, e.g., \emph{combination} because we wish binder units to exist not only for conjunctive (and-type) compositions of features and entities but also disjunctive (or-type) compositions along with anything in between the conjunctive and disjunctive extremes (i.e., implying the presence of multiple but not necessarily all components). As we will describe in Sec. \ref{subsection:WeightedDirectedLinks}, synapse weights make it possible to encode what type of composition (integration) one is dealing with along with the relative importance of each item in the integration.} it represents in the event that the integration occurs, and it also provides a mechanism for generating (potentially synchronized) activity in the component-representing neurons as a way of performing recall on partial cue. 

Because of their properties, which are consistent with the observed properties of some neurons as described in section \ref{ExpSupportNodes}, binder units can be interpreted as nodes in a semantic network standing for the integration they are able to recognize and recall. Wickelgren's \emph{chunk assemblies} \cite{Wickelgren92,Wickelgren99} are essentially the same as binder units, and Meyer and Damasio have suggested a theory of convergence-divergence zones for the brain, which also function in the same way \cite{MeyerDamasio2009}.

\begin{figure}
\centering
\includegraphics{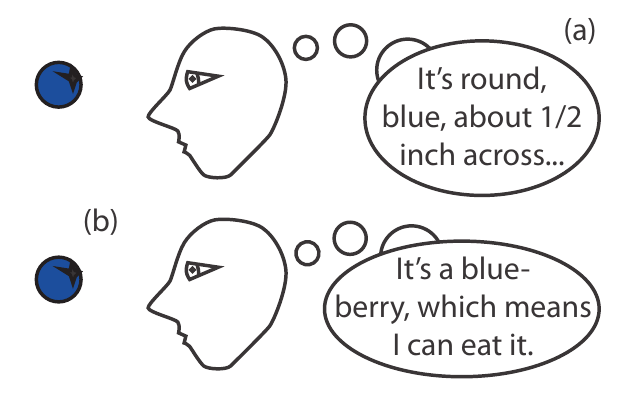}
\caption{Component (a) vs. integration (b) recognition. Recognizing a stimulus as an instance of a known integration is often at least as useful as recognizing its features.}
\label{fig:blueberry}
\end{figure}

In Fig.~\ref{fig:blueberry}, we show an organism (a person) encountering a stimulus, which is in fact a blueberry. At the component level, the person can see that the stimulus is blue, round and about half an inch across among other things. While this information is certainly relevant, it is at least as relevant and probably more so that the stimulus is like other stimuli the organism has encountered before and is indeed an instance of the type, `blueberry', informing the organism that the stimulus is safe to eat. Thus, binding encountered compositions into entities and types in memory which can be compared with future stimuli is very useful to an organism's function. In Appendix \ref{appendixBinderUnits}, we address the mathematical viability of experienced-based binder units.

\section{Framework} 

\label{Sec:Framework}

We suggest a neurally implementable semantic network methodology that is built around two foundational principles: that each node's meaning should be interpreted as a bound integration of the meanings of nodes (or external environmental events) that the node shares links with and that the network possesses a node for each discrete idea that it knows. We have seen considerable motivation for the first principle in the previous section on the basis of observations and theoretical considerations, and the second is one of our initial criteria. With regard to the latter, it is important to again note that the set of ideas known to a network---and therefore the nodes the network contains---need not be constant for advanced organisms. Nodes standing for very specific entities such as particular line segments in a visual stimulus may only be useful during presentation and processing of the stimulus, in which case they may be removed from the network after the stimulus is gone or shortly thereafter. More salient or frequently re-used nodes may be more long-lasting.

In this section, we begin by providing some additional motivation for the first principle and, in the process, clarifying it somewhat. Then we demonstrate the implications of our founding principles and how they allow various critical representation issues to be handled without violating our criteria for neural implementability (see pg.~\pageref{criterion:nodeMeaning}). Because the essentials of our method can be described using simple networks with unweighted and undirected links, this is how we will introduce them. We will use a sequence of examples of paradigmatic situations which will each demonstrate a representation issue along with the solution this framework offers. Afterward, we will show how weighted and directed links can be incorporated into the method in a way that expands its expressiveness.

\subsection{The bound integration interpretation}

In subsection \ref{subsection:Synch_Binding}, we pointed out the need for neural structures that serve to bind components together into recognizable aggregations. In our theory, this is precisely what nodes do with respect to adjacent nodes, i.e., those they share links with. There are a couple more reasons why this is an attractive interpretation for nodes in a neurally implementable semantic network.

One is that this interpretation ascribes the same basic semantic role to all links in a network with respect to the nodes they link together. Unlike nodes, which may share any number of links with other nodes, links themselves have but two attachments (the nodes they connect) along with weight and direction to determine their meaning. While ascribing a fundamentally different semantic interpretation to links on the basis of weight is possible, it would be strained. Moreover, this type of coding is not consistent with the role weights have in neural networks, which is about quantifying the level of influence neurons have on one another. It is better to have a conserved interpretation for links which is quantitatively \emph{modified} by weights---not changed.  For us, a link between nodes means the nodes are a component of each others meaning. The weight of the link indicates the level of this involvement. When we discuss weight and link direction further in subsection \ref{subsection:WeightedDirectedLinks}, we will motivate a distinction for the ``direction'' of involvement on its basis.

The third motivation we will mention for our interpretation is that neurons have the basic functionality of performing a thresholded, weighted sum of the activity among their inputs, which they pass along to their outputs. Units can therefore be seen as providing a summary of the varied activity among their inputs to their outputs. Conversely, because the units pass along their activity, in weighted fashion, to all their outputs, they provide a kind of proxy for this entire collection of outputs to each of their inputs. This aggregating faculty neurons fulfill with respect to connected units is kin to the semantic integration of linked neighbors into a single bound idea by nodes.

\subsection{Nodes as bound integrations}

\begin{figure}
\centering
\includegraphics[scale=0.5]{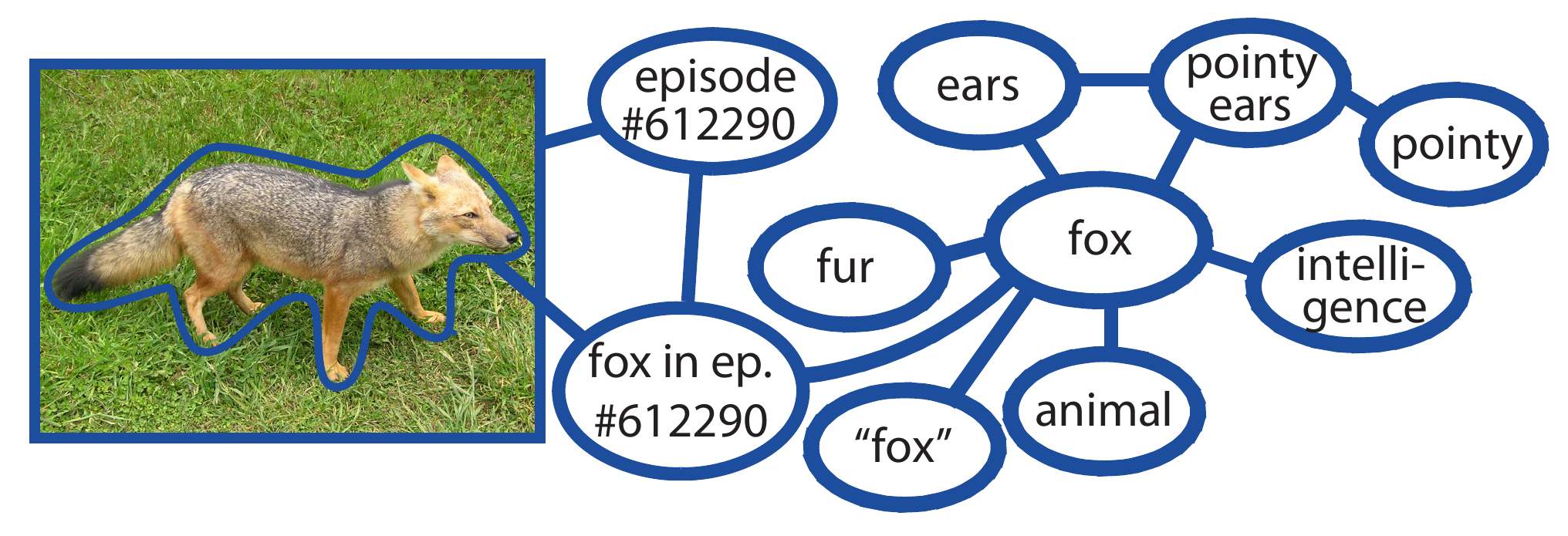}
\caption{Basic principle for node interpretation. Each node stands for a bound integration of the meanings of nodes and/or environmental events that it shares links with.}
\label{fig:foxdefnet}
\end{figure}

We now restrict to undirected, unweighted networks for the purpose of introducing important fundamentals that are independent of link direction and weight concerning the manner in which our proposed framework represents information. These networks may be thought of as a special case of directed, weighted networks in which all links are reciprocal and identically weighted.

We introduce the essential idea underlying our semantic network methodology with the example found in Fig..~\ref{fig:foxdefnet}. This figure contains a subnetwork that includes several nodes linking with and thus forming the integration that is the concept, `fox'. Included among these defining nodes are nodes standing for characteristics associated with foxes, such as `intelligence' and `pointy ears', along with a specific example of a fox. The diagram also includes a node for an episodic memory that involves that specific fox. The node for the episodic memory is defined by linkages with nodes that stand for the contents of the memory as well as with the node for the specific fox. The contents are represented by the image in the diagram, which is a shorthand for a set of nodes representing a specific visual stimulus. The specific fox node binds together the components of the visual stimulus corresponding to the fox and the fox concept into a single unit.

The meaning of all nodes in our framework is established according to this same basic paradigm. Each node links with other nodes that represent a component of its meaning. The node itself represents the integration, or coming together, of these components into a single unit. The bound integration interpretation for nodes relies on the presence of nodes in a network that link with events external to the network for the final grounding of meaning.  These events count directly toward the meaning of the linking nodes, which in turn provide the meaning for nodes linking with them, etc. Without the externally linking nodes, all nodes in a network would be meaningless.

\subsection{Avoiding repetition with instance and generic/type nodes} 

It is important to note that at the same time that the fox characteristic
nodes in Fig.~\ref{fig:foxdefnet} give definition to the fox concept, the fox concept gives
definition to the characteristic nodes as well by indicating a type of
entity that they commonly appear as part of. Similarly, while the `fox
in episode \#612290' node partially defines the fox
concept by providing an example of it, the fox node also partially defines
the `fox in episode \#612290' node by labeling the
entity it represents as a fox and thus establishing a connection
between that entity and the typical characteristics of foxes. In
this way, general and specific nodes make important reciprocal
contributions to each others' meaning.

\begin{figure}
\centering
\includegraphics[scale=0.5]{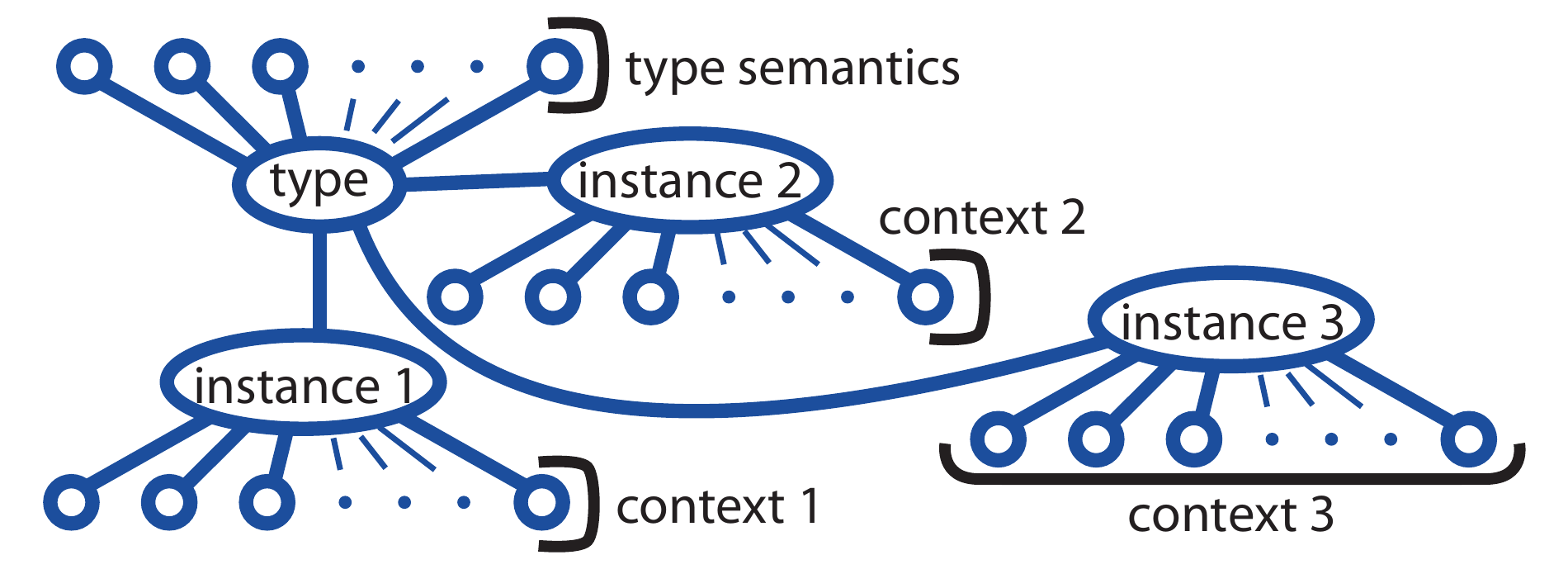}
\caption{Type and instance nodes. A frequently repeated semantics can be attached to many instances by using a type node, which links both to the nodes necessary to encode that semantics and to the multiple instances of the semantics, which acquire their own special meaning by way of connections to nodes expressing their context, other types they belong to, etc.}
\label{fig:typesandinstances}
\end{figure}

The principle we have just described is what permits our semantic network methodology to place many nodes in a semantic network whose meanings have much in common without violating our efficiency criterion: to avoid excessively repetitive encoding (see pg.~\pageref{criterion:DRY}). The similarities in meaning for such nodes, e.g., common fox characteristics if the nodes stand for individual foxes, link with a single generic or type node. The similarities are then acquired by the individuals, i.e., instances of the type, via a link between the type node and every instance node. We illustrate this principle in Fig.~\ref{fig:typesandinstances}. For $n$ instances of a type with $m$ associated traits, only $n + m$ links are needed rather than $n \cdot m$ as would be required if the traits were encoded independently on each instance node.

Instance nodes acquire specific meaning via links to nodes indicating their specific context. They can also link to any characteristics specific to themselves and any exceptions to the traits of types they are associated
with\footnote{M. Ross Quillian and Allan Collins have suggested this
  model for memory organization previously and have provided evidence
  to support it: in reaction time experiments, it takes less time for
  human participants to accurately confirm the special traits of entities or concepts belonging to a
  given class than to confirm traits typical of the class to which they belong \cite{CollQuill1969,CollinsLoftus}. For instance, it takes less time for participants to confirm that canaries can sing than that they have skin. This indicates that an entity's or concept's specific traits are stored in closer proximity to the item itself than are the traits of classes the item belongs to.}.
Instance nodes can be subtypes with links to their own instances.

\subsection{Representing relational structure with type and instance nodes}
\label{Rel&RelPosTypeNodes}

\begin{figure}
\centering
\includegraphics[scale=0.5]{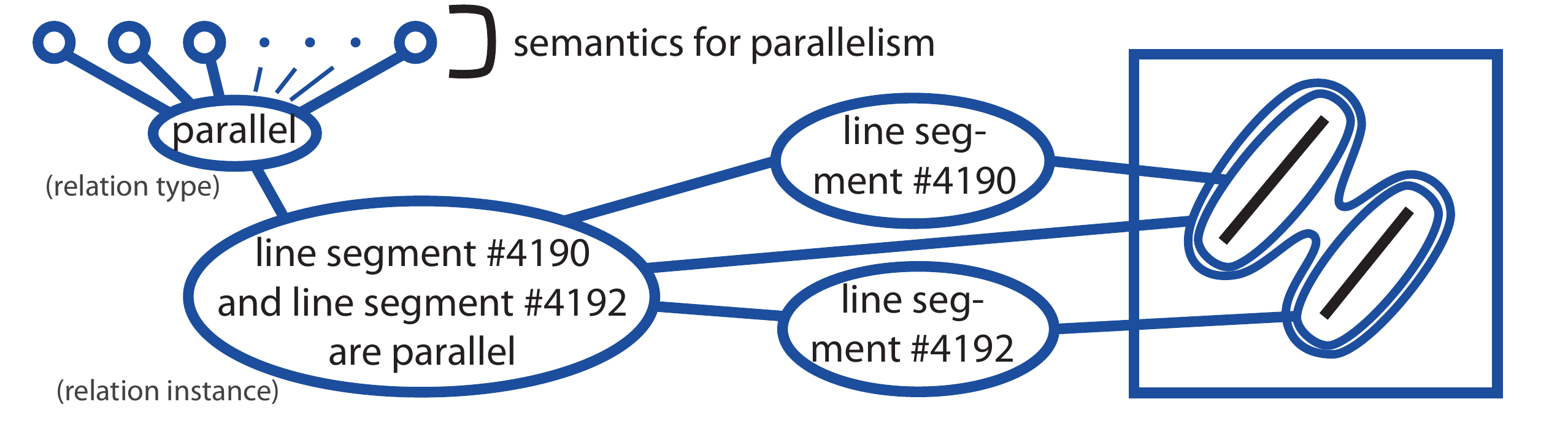}
\caption{Relational types and instances. The type-instance method is used in the context of relationships to place a gating node between two other nodes that stands for the relationship that the two nodes have. The semantics of the relationship is acquired indirectly by linking the gating node (relationship instance node) to a node standing for the relationship type.}
\label{fig:relationTypeInstance}
\end{figure}

Type-instance machinery provides the ground for our treatment of neurally implementable relational structure encoding. A relationally structured idea, such as ``line segment \#4190 and line segment \#4192 are parallel'' as shown in Fig.~\ref{fig:relationTypeInstance} involves multiple components which are related to each other in one or more ways. In this case, the components are the two line segments, and parallelism is the single relationship between them. This specific case of parallelism between these specific line segments can be treated as an instance of a type: a relationship type, parallelism---because this will not be the only example stored in the network of line segments being related in this way. Specific line segments are the context for a specific case of parallelism. A node standing for a specific case can therefore establish its meaning by a link with a type node for `parallel' and by links with nodes representing the line segments.

Relationship type and instance nodes such as these are the first of two major type-instance devices used to encode relational structure in our method. The second major element, relational position type and instance nodes, will be explained very shortly. Types for relationships such as `parallelism', `bigger-than' and `ownership' should have nodes in a semantic network that knows them, since they are semantically discrete ideas. Specific instances of these, such as ``A house is bigger than a car,'' should also have nodes for the same reason. Like all type nodes, the meaning of a relationship type node will be expressed by the node's linkages with nodes standing for characteristics of the relationship and to examples of the relationship in specific contexts. The specific instances of the relationship are expressed as involving this common meaning by way of links between the instance nodes and the relationship type node. A structured situation will be represented by including a relationship instance node for each of the relationships involved in it.

\begin{figure}
\centering
\includegraphics[scale=0.5]{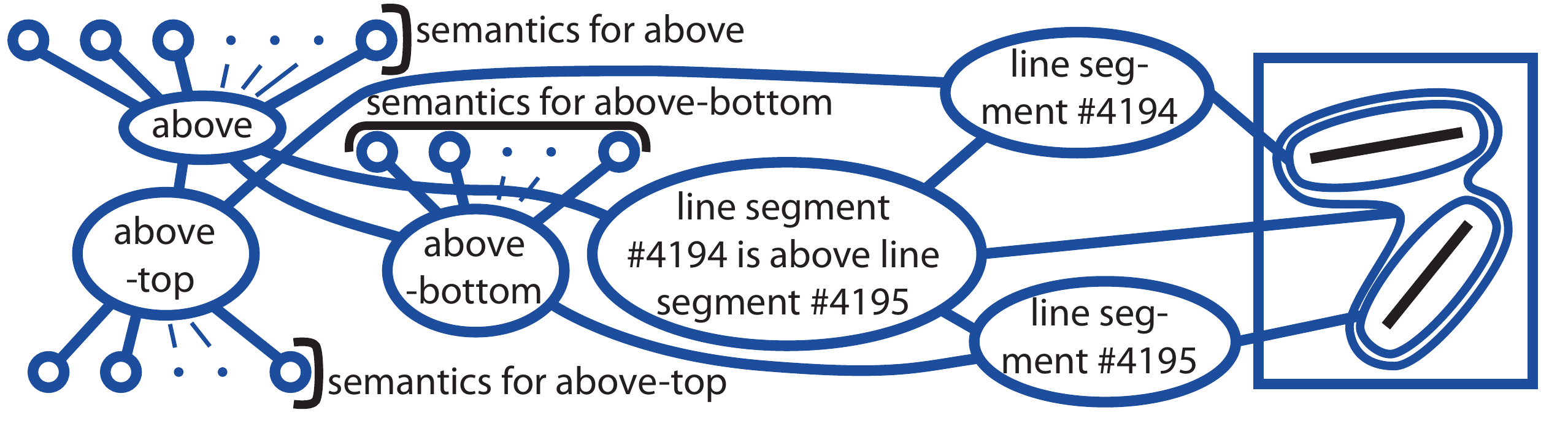}
\caption{Encoding relationships that have internal structure. Type nodes for relational positions are used to designate the special position each node occupies in the relationship.}
\label{fig:relationInternStruct}
\end{figure}

In most cases, it will be necessary to do more than this to express
relational structure, however. This is because most relationships
involve special positions occupied by ideas occurring in the
relationship. The special positions are typically different from each
other in important ways. For example, in
Fig.~\ref{fig:relationInternStruct}, the relation `above' involves a
`top' and a `bottom', which are fundamentally different. The same will be
true for a concept corresponding to any transitive verb, which will
involve special roles for the direct subject and object. It will also be true for spatial
relationships, temporal relationships, sequential relationships and generally
any relationship that involves contrast.

The unique positions involved in such relationships are discrete ideas and amount to a significant component of the relationships' meaning, which indicates they should have nodes. Relational position nodes also make it possible, when defining relationship instances, to specify which of the ideas participating in the relationship occupy which roles. In our method, explicit nodes standing for relationship positions link with the relationship type they participate in (because they contribute to the relationship's definition and the relationship contributes to theirs), with any other nodes giving definition to the relational positions, and with nodes implementing the relational position in the context of specific instances of the relationship. The relational position nodes are type nodes.

Relational position nodes are demonstrated in Fig.~\ref{fig:relationInternStruct}. `Above' is the relationship type, and as we have said it involves two positions: `above-top' and `above-bottom'. The relationship instance is represented as such in the way we have already described, viz., by linking with the relationship type, and the line segment on top is represented as being on top by a link between its node and the `above-top' position and similarly for the segment on bottom.

There is a connection between the method we are presently describing for encoding relational position and what is known as \emph{keyword coding} in computer programming. Programmed procedures, functions and commands usually involve parameters with specific roles that determine what happens when the function is called. For example, a \emph{Draw()} function, when called, may accept the parameters, \emph{shape}, \emph{size} and \emph{color}, and produce a corresponding figure on a computer screen. These unique parameters are analogous to the unique roles played by various ideas in a relationship because the values passed into the different parameters, like the ideas occupying the different roles, are not all treated the same. A relation itself can be considered a function that receives ideas as parameters associated with specific relational positions and produces a meaning as output.

Differing parameters of a function must be somehow distinguished from each other when they are passed as input to the function in order for them to be handled properly. Classical coding methods typically associate a specific role with each position in the ordered list of parameters occurring in the function call. Thus, if the correct order for \emph{Draw()} is  \emph{shape}, \emph{size}, \emph{color}, then \emph{Draw(`circle', 100, `sky blue')} will produce a light blue circle 100 pixels in diameter whereas \emph{Draw(100, `circle', `sky blue')} will produce an error. Doing things this way requires both the user making the function call and the computer executing the function to keep track of position in an ordered list, which is usually a fine approach for simple functions. However, in the case of complex functions with many adjustable parameters, the majority if not all of which have default values that the user may not want to change, it becomes prohibitive to ask the user to recall the exact position at which all the various parameters are supposed to occur every time the function is used. In this case, keyword coding is much more advantageous.

Keyword coding attaches a keyword to each parameter value occurring in the function call. The keyword names names the parameter that the value should be passed into. For example, either \emph{Draw(shape = `circle', size = 100, color = `sky blue')} or \emph{Draw(size=100, shape = `circle', color = `sky blue')} will produce the correct result under such a scheme. In this way the parameter is specified explicitly rather than via an inference based on its position in a list.

The relevance here is that, for neurally implementable semantic networks, differentiating roles among nodes involved in a relationship purely on the basis of the weights and directions of these nodes' links with the relationship instance node, while conceivable, would be strained and disconnected from biological reality in which neurons, to good approximation, simply add up their weighted inputs indiscriminately, cf. the single- and two-compartment neuron models \cite{HerzEtAl}. For example, an order encoding for relational role specification, such as the subject-verb-object order used in English language, seems untenable for us because it appears difficult to specify which, among synaptically implementable links between the relationship node and the related nodes, should be considered `first', `second', or `third'. The strategy suggested here of specifying relational position by linking ideas participating in a relationship to relational position type nodes is analogous to attaching a parameter-specifying keyword to data passed to a function.

\subsection{Frequent need for separate relational position instance nodes}

\begin{figure}
\centering
\includegraphics[scale=0.5]{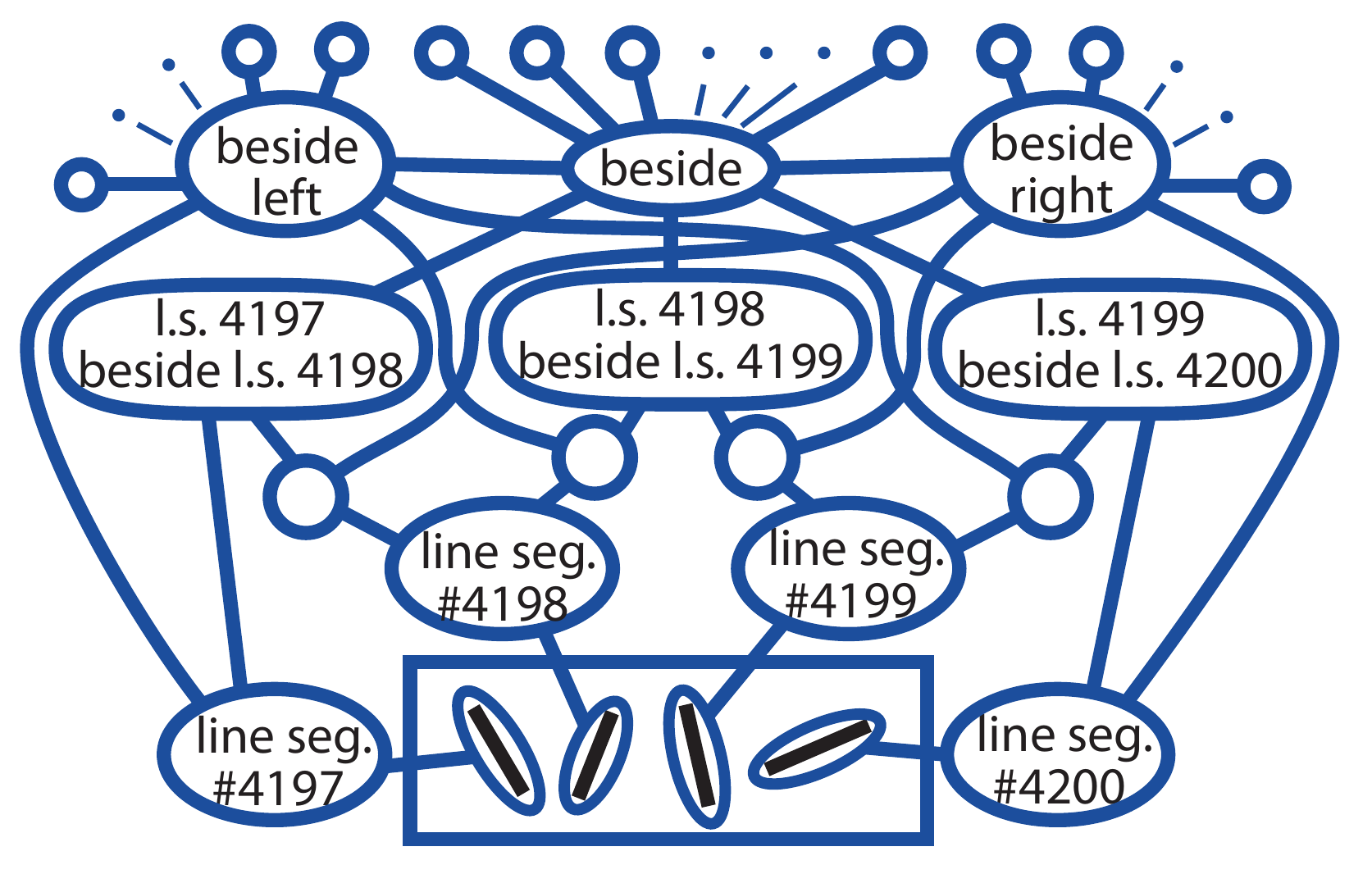}
\caption{The use of independent relational position instance nodes in the frequent situation that an idea occupies different relational positions in the context of different relationships. The four larger unlabeled nodes stand for relational position instances. They are needed here as distinct nodes because the `line segment \#4198' and `line segment \#4199' are alternatively a `beside-left' and a `beside-right' depending on context.}
\label{fig:multRelInst}
\end{figure}

Another crucial and essential principle is yet required in order to represent most cases of relational structure. The need for it is illustrated in Fig.~\ref{fig:multRelInst} in which there are three relationships represented, each indicating that two of the four line segments shown are beside one another. The `beside' relation involves two positions, `beside-left' and `beside-right.' The two middle line segments each occupy both the -left and the -right positions in the context of one of the two instances of `beside' that they participate in. The goal is to specify, for both segments, in which relationship instance the segment occurs on the left and in which it occurs on the right. 

This cannot be done by simply linking the nodes for the segments directly to the type nodes for the positions. Taking such an approach would result in the nodes for the middle two line segments each having a pair of links with both relational position types. While this specifies that each segment is on the left of something and on the right of something else, it does not clarify in \emph{which} relationship the segment is on the left versus on the right.

Generally speaking a single idea may, and frequently will, participate in multiple relationships and therefore occupy multiple relationship positions. Often, the position held by the idea in one relationship is very different from or, as in this case, completely opposite to the position it holds in another relationship. When this happens, links between the idea and the various positions it occupies don't clarify which position the idea holds in the context of each individual relationship.

We remedy this by including independent nodes standing for the relational position instances (these are the larger unlabeled nodes in the figure). The relation instance to which each relational position instance belongs is clear from the position instance's link with this relation instance and no other. The ideas occupying the position instances (the line segments in this case) are then easily represented as such by way of a link between the idea and the position instance.

An important clarification to make is that independent relational position instance nodes are not always to be used. Like all nodes, they are to be used only when a separate bound idea is necessary in order to convey important or desired meaning. In the case that a particular idea only occupies a single relational position in a single relationship instance, it is fine (and preferred) to omit the separate relational position instance node and link the idea's node directly to the correct relational position type node and relational instance node. In effect, this can be thought of as a merging of the idea and the relational position instance into a single bound idea. 

\subsection{Demonstration of undirected, unweighted case}
\label{sec:DemoUndir}

\begin{figure}
\centering
\includegraphics{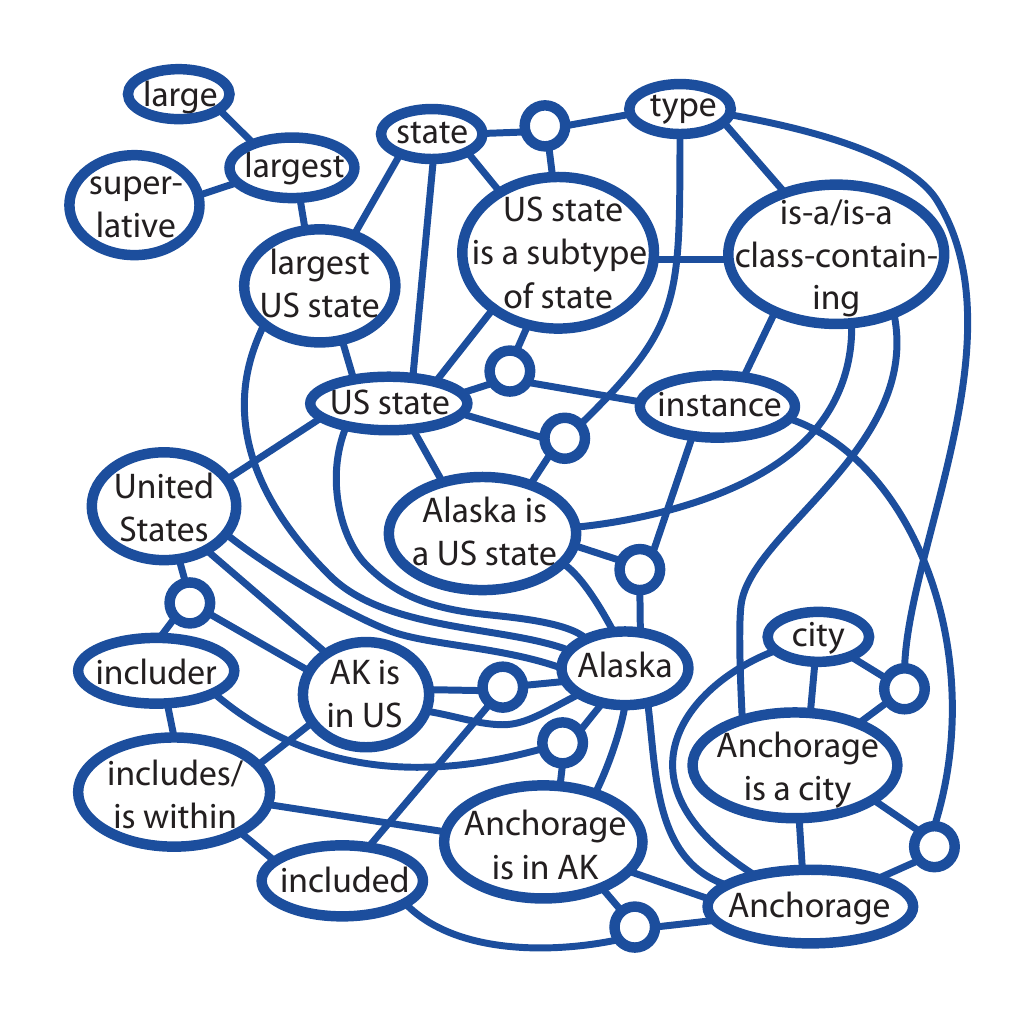}
\caption{Semantic network utilizing our schema and expressing knowledge related to Alaska. The small unlabelled nodes correspond to relational position instances as can be inferred from their connections to both relational position types and specific contexts.}
\label{fig:SemNet_undirected}
\end{figure}

The principles we have introduced so far are sufficient to construct simple unweighted, undirected networks expressing relationally structured knowledge. The network in Fig.~\ref{fig:SemNet_undirected} represents some basic facts pertaining to Alaska, such as that it is a US state, that it is the largest among US states, and that Anchorage is a city it contains. This network uses two of the most essential semantic relations, inclusion and subsumption (is-a). The direct link between `state' and `US state' allows `US state' to access the semantics (not shown) associated with the type, state. To explicitly represent the fact that these two ideas, `state' and `US state', are participating in an `is-a' relationship, the network includes a ``US state is a subtype of state'' relationship instance node and the associated network structure in the top right-hand corner of the figure, which places the ``US state is a subtype of state'' node as an instance of the ``is-a/is-a-class-containing'' relationship type and connects `state' and `US state' with their correct positions in the relationship. In precisely the same manner, the direct link between the `United States' and `Alaska' nodes expresses that these two ideas are involved with one another, and the specific nature of that involvement, i.e., their relationship, is expressed by the ``AK is in US'' node and associated structure in the bottom left-hand corner.

\subsection{Extension to weighted and directed links}

\label{subsection:WeightedDirectedLinks}

As we have said before, link weights allow a network to specify \emph{how} involved one idea is in another rather than just \emph{that} the two ideas are involved. For example, flowers and petals are involved concepts because petals are very important parts of flowers. So are bees and flowers since bees pollinate flowers, and flowers feed bees. But clearly there is a difference in the level of involvement here. Insects other than bees can pollinate flowers, and pollination is only one part of a flower's life whereas petals are common to all flowers and are one of their most distinctive features. Link weights allow this to be expressed. Greater weight means greater involvedness or \emph{criteriality} \cite{CollinsLoftus}.

Thinking again about the neural implementation of nodes, recall that there will be some activity associated with a node indicating its level of presence in mind. This activity will be implemented on neural action potentials. Neurons, to a good approximation, perform a weighted sum on the activity in neurons connecting to them in determining their own activity, which will be an increasing function of this weighted sum. Weights among inwardly directed links thus correspond to the level of influence a linking node has on the receiving node's activity. This is consistent with the criteriality interpretation for link weights because it means the link weight determines the degree to which the presence of the idea corresponding to the linking node determines the level of presence for the idea represented by the receiving node. Activity in a `bee' node, for example, will not have as much impact on activity in a `flower' node as will activity in a `petal' node if the incoming weights to the `flower' node reflect the relative importance of `bee' and `petal' to `flower'.

The outgoing weights should reflect relative importance as well. For example, activity in a `flower' node should impact activity in a `petal' node more strongly than it should activity in a `bee' node. This is because the presence of a flower almost certainly indicates the presence of petals whereas it only indicates the \emph{possibility} of one or more bees. Therefore, activity in the `flower' node for a network should, in most cases, propagate strongly to the network's `petal' node whereas it should only propagate strongly to the `bee' node in the event of facilitating activity in nodes representing other phenomena indicative of bees, such as the familiar buzzing sound made by winged insects. Weighting the link from `flower' to `petal' more strongly than the weight from `flower' to `bee' is consistent with this desired dynamics. In general, we expect a central node's activity to impact the activity of the nodes it links with according to how criterial each of these nodes is to the central node's meaning. Outgoing weights should therefore also reflect criteriality.

\begin{figure}
\centering
\includegraphics{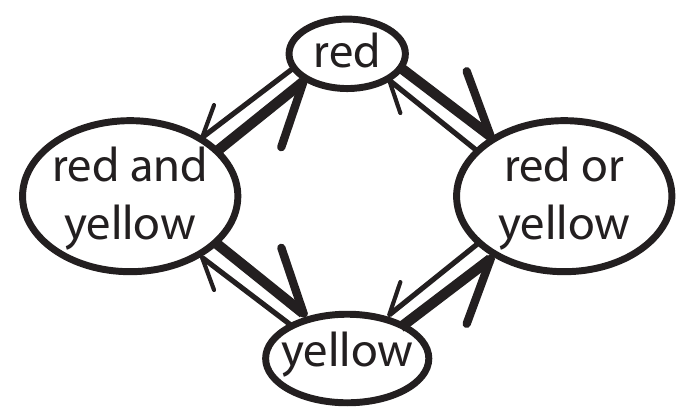}
\caption{Different incoming versus outgoing weight biases for `and' and `or' compositions.}
\label{fig:redAndOrYellow}
\end{figure}

\newcounter{footnote:Ior}[footnote]
\addtocounter{footnote:Ior}{1}

It will be common, however, that outgoing weights should be somewhat to substantially different from incoming weights on the basis of dynamical considerations. Consider, for example, an `or' composition versus an `and' composition such as `red or yellow' versus `red and yellow' as shown in Fig.~\ref{fig:redAndOrYellow}. The `and' requires the presence of all composing entities or concepts whereas an `or' requires the presence of only one. In this case, the `red and yellow' composition could be satisfied by red and yellow stripes, yellow dots on a red background or vice versa, as well as many other possibilities involving appreciable amounts of both colors. `Red or yellow' will be satisfied by each of these as well as conditions involving only appreciable amounts of one of the colors\footnote{We are using inclusive `or' here.}.  Therefore, we expect the incoming weights from the composing nodes (viz., the `red' and `yellow' nodes) to be larger in the case of the `or' composition than they are the case of the `and' composition since activity in only one of the composing nodes should be sufficient for transmission in the `or' case whereas activity in both should be necessary for transmission in the `and' case.

On the other hand, we expect precisely the inverse discrepancy for the outgoing weights. `Red and yellow' means that both `red' and `yellow' are present whereas `red or yellow' means that either or both could be present, but it's unclear which without further information. The `and' composition should indicate the definite presence of all composing elements whereas the `or' should indicate possibility only for each of them specifically. Therefore the outgoing weights from the `and' composition should all be sufficient to transmit activity whereas those from the `or' composition should be considerably weaker.

This `and' versus `or' example introduces the two types of meaning that are associated with the two different directions for links in our methodology. Outgoing weights depict what we shall call \emph{consequential} meaning, and incoming weights depict \emph{conditional} meaning. Conditional meaning refers to the bundle of ideas constituting \emph{conditions} under which a given idea may be said to be present whereas consequential meaning refers to the bundle of ideas that may occur as a \emph{consequence} of a particular idea. We have seen, in the case of both `and' and `or' compositions, that criteriality for constituent ideas in a bound integration can be different for these two types of meaning. As another example, consider the idea, `rain'. In the conditional sense, rain means that drops of water are falling from clouds in the sky. In the consequential sense, rain also means moist soil, puddles, poor traction and possibly wet clothes (or fur). The conditional and consequential meanings are clearly different in this case as well, and both meanings are relevant to an organism's concerns.

An unequivocal demonstration that conditional and consequential meaning are distinct yet both important can be found at the extreme sensory and motor ends of the central nervous system. A node corresponding to a particular muscle movement will have, as its consequential meaning, the muscle movement that it causes. Such a node's conditional meaning will be all the different actions which the muscle movement could be involved in along with any particular situations or sensory events for which that exact muscle movement is a good response. On the other side of the nervous system, a node standing for a particular environmental stimulus, such as light of a given frequency range falling on a certain region of the retina, will have the conditional meaning that this environmental event has taken place. The consequential meaning for such a node will include all the various objects, color patterns, etc., that this event could be a component of. In both cases, conditional meaning and consequential meaning are quite different and also quite relevant.  In general, these two types of meaning will be overlapping but not strictly the same, e.g., in the absence of direct evidence, puddles, poor traction and wet clothes can be indicative of rain.

\begin{figure}[htp]
\centering
\includegraphics{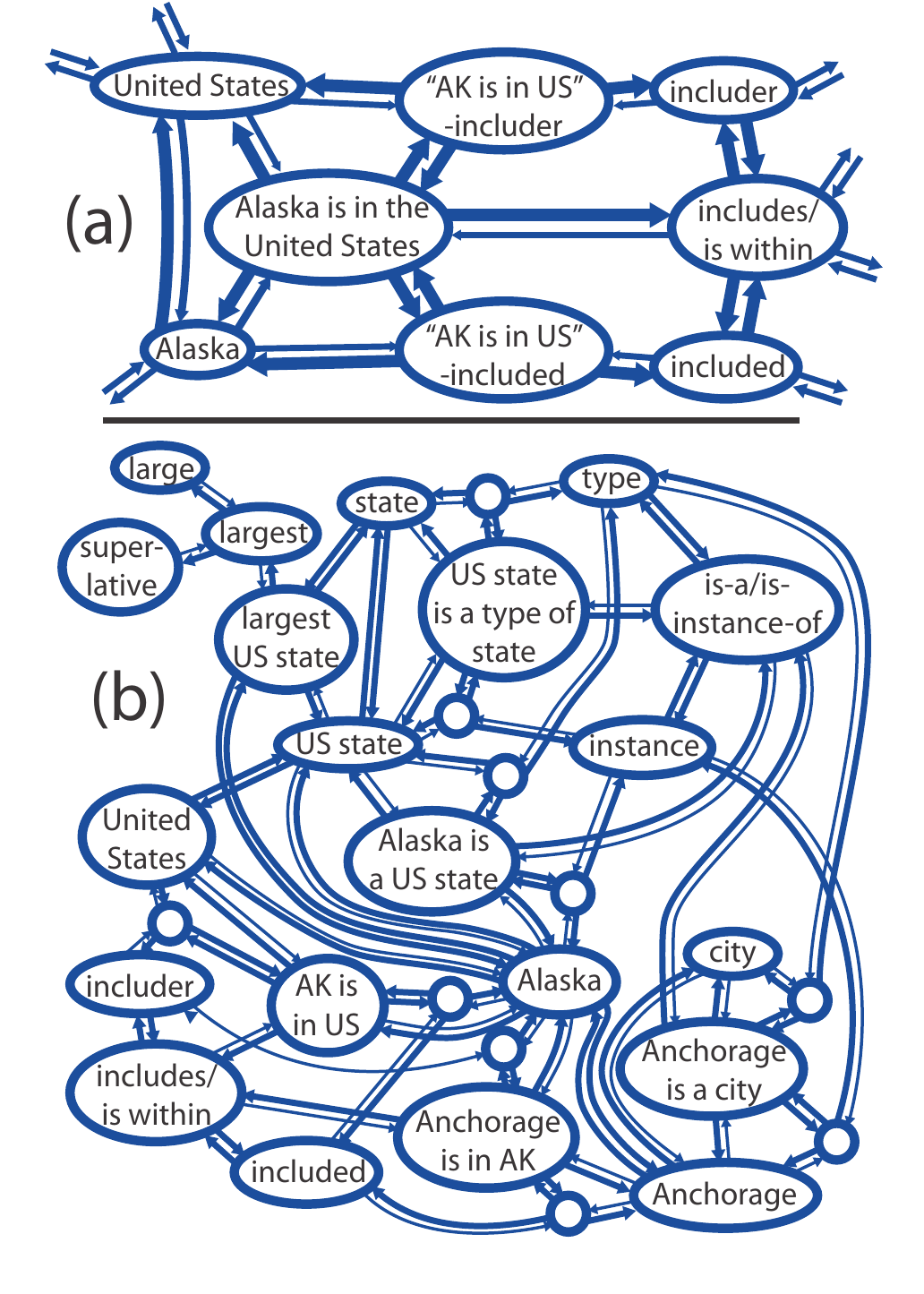}
\caption{Directed, weighted version of the Alaska network in Fig.~\ref{fig:SemNet_undirected}. Weights reflect criteriality, and direction specifies conditional versus consequential meaning (see text). In (a) a sub-network encoding the relationship, ``Alaska is in the United States'' is magnified. Here and in general, more abstract nodes tend to receive stronger weights from more specific nodes. This indicates that the specific nodes participate more strongly in the abstract nodes' conditional meaning, and the abstract nodes participate more strongly in the specific nodes' consequential meaning than vice versa. In (b), the extended network is shown.}
\label{fig:dwAlaskanet}
\end{figure}

\subsection{Demonstration of weighted, directed semantic networks}

We have now provided sufficient background to present a directed, weighted version of the Alaska network from Sec. \ref{sec:DemoUndir}, which we show in Fig.~\ref{fig:dwAlaskanet}. In part (a) of this figure, we magnify a section of the network which expresses the relationship, ``Alaska is in the United States,'' in order to shed light on the directional weight asymmetries occurring within relationship-specifying sub-networks. For example, the weight from the ``Alaska is in the United States'' node to the `includes/is within' node is stronger than vice versa. Similarly, the weight from ```AK is in US'-includer'' to both `includer' and `United States' is stronger than the weights for links in the opposite direction. This is because activity in a more specific node should usually produce activity in the more general nodes it is associated with whereas activity in a general node should only produce activity in the specific nodes most relevant to the present context if any. Therefore, more general nodes receive stronger links from more specific nodes than vice versa.

\begin{figure}[htp]
\centering
\includegraphics{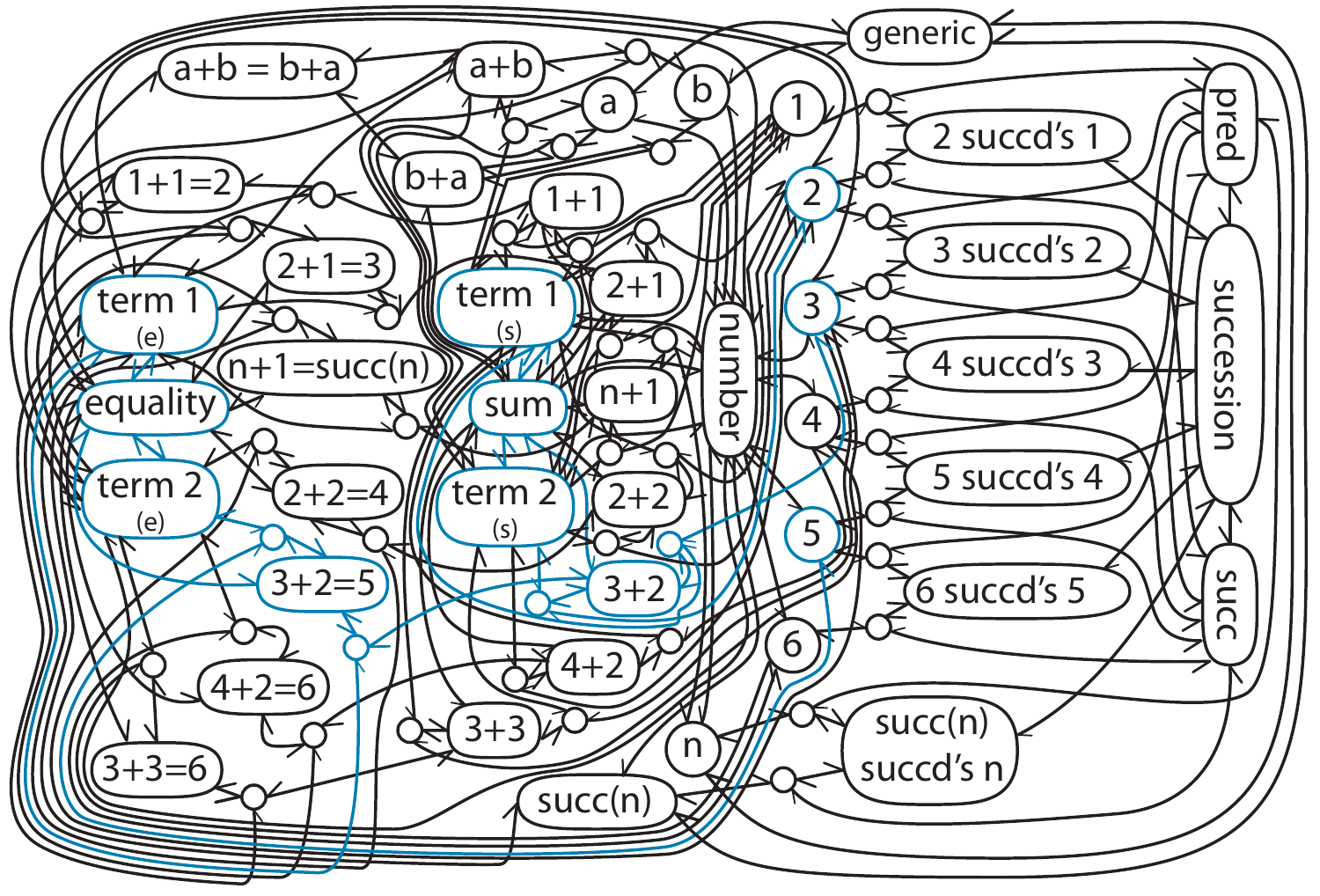}
\caption{Network describing the numbers one through six and some important relationships. Weights are indicated by arrowhead sizes. The knowledge encoded here might be similar to that of a child who is first learning numbers. While the child likely would not know all the words used here in the node labels, the concepts the child uses could be similar without necessarily being associated with specific names. Highlighted pathways and nodes show the portion of the network most pertinent to `3 + 2 = 5'.}
\label{fig:numberNetHL}
\end{figure}

Finally, to illustrate the performance of our semantic network schema in the context of a familiar and fairly structured domain, we provide a network depiction of the numbers one through six and some relationships in Fig.~\ref{fig:numberNetHL}. At the core of this network are several instances of the `succeeds' relationship which occur between consecutive pairs of the numbers and which indicate the basic structure among them. The general concept of `number' is defined by a `number' node linking with each of the specific numbers as well as with other items in the network which will be numbers, such as the terms in the `sum' concept and the `equality' relationship. The type, `sum', is encoded as a composite of two terms, both of which are numbers. This type is defined by way of links with stored instances of `sum', including `1+1' and `3+2.' These, in turn, are represented as having something fundamental in common by way of their common link to the `sum' type. The same is done with `term 1 (s)' and `term 2 (s)', the types for the components of `sum'. Each is linked with an instance of itself as found in a particular sum. This defines the types by way of examples and establishes the linked instances as being involved in the sum they belong to in a similar way.

The same type-instance machinery is applied to the equality relationship and the positions within it as well. Instances of the equality relationship provide representation of arithmetic facts, e.g., `2+2=4', and commutativity of addition, `a+b=b+a'. Storing the commutativity relation allows the network to do without explicitly storing both orderings of asymmetric sums, such as `3+2=5' and `2+3=5': The latter may be deduced from the former along with commutativity of summation. Nodes and pathways central to the relationship, `3+2=5', are highlighted so that the reader can verify that this relationship and its internal construct, `3+2', are encoded according to the principles we have outlined. The network contains 88 nodes and 366 directed links. For a complete textual representation of the network in Fig.~\ref{fig:numberNetHL}, please see Appendix \ref{TextNumberNet}.

While this network represents only the numbers one through six, we can imagine that in a network representing numbers in general, not all numbers and numerical facts need to have a permanent representation---only those sufficient for deducing important numerical facts when needed. In an Arabic number system, for example, arithmetic facts need only be stored for pairs of single digit numbers. Rules prescribing how these facts may be used to deduce relationships involving large multi-digit numbers (which the individual may have never even encountered before and never will again) then make it possible to obtain such relationships reasonably quickly without having to store them permanently. While the network in Fig.~\ref{fig:numberNetHL} is clearly a truncation of a fully developed representation of numbers, one can envision how the basic principle illustrated can be extended to include all necessary single digit arithmetic identities and deductive rules necessary to form a mature working knowledge of numbers and basic arithmetic.

We also note that the network in Fig.~\ref{fig:numberNetHL} is incomplete because it does not ground itself in nodes that link to the environment. Because our purpose here is only to provide a portion of a network encoding basic relationships involving numbers, we have left these out. We can envision this grounding beginning to occur, however, via links between the individual numbers 1-6 and nodes for previously encountered sets of entities containing the corresponding number of entities. For example, `1' might be linked with the sun or a single apple whereas `2' might be linked with a pair of shoes, `3' with the vertices of a triangle, and so on. Additionally, the `sum' concept may be linked with a node standing for the physical act of combining two sets of entities and `equality' with comparing two sets and verifying that they contain the same number of entities. `Succession' also may receive further grounding through specific stored examples of consecutively occurring entities in a sequence, such as notes or words in a song or events in a story or episodic memory.

To assess the efficiency of our semantic network method, let us examine the information content of the network in Fig.~\ref{fig:numberNetHL}. The information content of a directed network may be defined as the amount of information necessary to specify the number of nodes in the network out of some possible number plus the amount necessary to specify the connections among the nodes:
\begin{equation}
h_{\rm net} = h_{\rm size} + h_{\rm connections}.
\end{equation}
The first amount, in bits, is the base 2 logarithm of the maximum network size, which for our purposes will be certainly no greater than $10^{16}$, the number of \emph{synapses} in a human brain. This yields, at the very most, 53 bits for $h_{\rm size}$, which we will see is small compared to $h_{\rm connections}$ and will be neglected heretofore in our expression for $h_{\rm net}$.

The amount of information in the connections is equal to the amount of information necessary to specify the network's wiring diagram (which units are connected to which) plus the amount necessary for specifying the weights of the connections:
\begin{equation}
 h_{\rm connections} = h_{\rm wiring} + h_{\rm weights} .
\end{equation}
Supposing the density of connections to be relatively fixed, $h_{\rm wiring}$, in bits, is given by
\begin{equation}
h_{\rm wiring} = \log_2\left( {\begin{array}{*{20}c} {n \cdot \left(n-1\right)} \\ l  \\ \end{array}} \right),
\end{equation}
where $n$ is the number of nodes and $l$ the number of directed links in the network.

Using Stirling's approximation,
\begin{equation}
h_{\rm wiring} = n\cdot \left(n-1\right) \cdot \log_2 \left( \frac{n\cdot \left(n-1\right)}{n\cdot \left(n-1\right)-l} \right) + l \cdot \log_2 \left( \frac{n\cdot \left(n-1\right)-l}{l} \right).
\end{equation}

Since $l$ is small compared with $n^2$ and $n$ is large, we may further approximate:
\begin{equation}
h_{\rm wiring} = n^2 \cdot \frac{l}{n^2} \cdot \log_2 \left( e \right) + l \cdot \log_2 \left( \frac{n^2}{l} \right) = l \cdot \log_2 \left( e \cdot \frac{n^2}{l} \right).
\end{equation}
The information in the weights is simply:
\begin{equation}
 h_{\rm weights} = l \cdot h_{\rm link} ,
\end{equation}
where $h_{\rm link}$ is the amount of information necessary to specify a single link weight. Supposing 16 distinguishable levels for the weight gives $h_{\rm link} = 4$ bits.

The total network information in bits is then approximately:
\begin{equation}
h_{\rm net} = \left( 4 + \log_2 \left( e \cdot \frac{n^2}{l} \right)\right) \cdot l .
\end{equation}
This gives about 10 bits per link for networks with similar sparsity to the network in Fig.~\ref{fig:numberNetHL}, for which $h_{\rm net} \approx 3600$ bits. While the network may seem complex, this amount is comparable to the information content of a text rendering, such as that found in Appendix~\ref{TextNumberNet}, for the same set of facts. The text in Appendix~\ref{TextNumberNet} contains about 1600 characters, requiring 13,000 bits under standard 8 bit per character encoding. Applying the gzip compression algorithm to a file containing the text gives a 4200 bit file. This is remarkably close to the number of bits in our network representation.

The amount of information in Fig.~\ref{fig:numberNetHL} is also tiny compared to the information capacity of the brain, which, in bits, is on the order of the number of synapses, $10^{16}$---about a billion times more.

\section{Discussion}

One dimension of neural network structure we have not tried to incorporate into semantic network structure in the present work is inhibitory or negatively weighted connections. There are reasons for this. First, neurons generally make either strictly excitatory or strictly inhibitory connections to other neurons, not a mixture of both. This structural property poses obstacles for the neural implementation of a semantic network structure in which linkages can be either positive or negative. One possibility is for inhibitory neurons to be incorporated into nodes themselves, facilitating direct inhibitory connections from neurons associated with one node to neurons associated with another. The difficulty with this is that any inhibitory neurons involved in a node would be unable to make any excitatory connections onto other neurons in the node so as to facilitate the node's structural coherence and ability to function as a single unit. Another possibility is to implement inhibitory links between nodes with targeted disynaptic pathways. Such pathways present an order of complexity that, while possible, we would prefer to postpone for the present. 

There are also experimental results casting doubt on node participation by inhibitory neurons. During observations of the songbird forebrain nucleus HVC, which as we have previously mentioned is involved in both producing and recognizing song and which contains excitatory neurons that fire in conjunction with specific times during both the heard and vocalized song, inhibitory interneurons are observed to fire throughout the song \cite{HahnloserKozhevnikovFee}.  While these observations are consistent with the organization of excitatory neurons in this nucleus into cell assemblies representing specific points in time during the song, they are inconsistent with the incorporation of the inhibitory interneurons into those assemblies, indicating a less specific role for inhibitory neurons. Possibilities in this regard have been demonstrated by computer models, cf.~\cite{ChangJin} in the case of HVC itself and \cite{BushDouglas} in the case of cortex. On the other hand, there is consistent temporal structure in the observed inhibitory interneuron activity in HVC during singing including robust silent periods, so disynaptic inhibitory pathways between nodes should not be ruled out on the basis of these experimental results.

For these reasons and for the sake of simplicity, we have endeavored here to examine the semantic representation capacities for a network of positively weighted links only. Inhibitory neurons can still play a role in implementing the dynamics of such a network, for example, by limiting the number of simultaneously active nodes. We will comment that the extension of our semantic network schema to one including negative weights between nodes is fairly straightforward: nodes connected by negatively weighted links should be interpreted as specifically excluded from the bound integrations constituting their respective meanings. Negatively weighted links would aid in the representation of relationships involving negation such as ``The chair is not green.''

The neurally implementable semantic network schema we present brings together known results regarding neural response to abstract concepts, semantic network theory, cognitive architectures, cell assembly theory and basic principles of neural interconnection and dynamics. The result is a framework that offers a starting point for future work concerning meaning and symbolism in neural networks. The central principle that our schema is built around is that of interpreting neurally implemented nodes as bound integrations of the meanings of nodes they share synaptically implemented links with as well as environmental events they either cause or are caused by. We have shown how this basis makes it possible, through the use of relation and relational position type and instance nodes, for relationally structured information to be represented in a semantic network without invoking any elements that would be especially unnatural to manifest neurally.

It remains to be seen, however, whether meaning can truly be stored by neural networks in the way that we suggest. Two levels of theoretical development and testing may be helpful in making this determination---the dynamic semantic network level and the neural network level. By the dynamic semantic network level, we mean a theory ascribing activity to nodes in a network and having principles that describe both the flow of activity among the nodes as well as structural changes to the network itself, such as the addition of new nodes and links and the adjustment of link weights. A dynamic semantic network theory of this kind will need to be based on experimentally observed principles for neural dynamics, and it will need to be able to store and recover information in the semantic network's structure via these dynamics.

We have mentioned the role that temporally correlated, e.g., synchronous, activity seems to play in binding various components of a current stimulus together as entities \cite{Singer1999}. Our model for neurally implementable semantic representation focuses on bound integration among \emph{stored} mental components. For a dynamic model, binding among \emph{active} components is necessary also, and it could play a role in deciding which nodes should become bound together via common links with a new binder node. It may therefore be worthwhile or even necessary to have a model for activity on semantic networks that includes active binding somehow---possibly by using some type of network model for activity itself with links representing temporally correlated spiking.

If a dynamical framework at the semantic network level is successfully proposed and validated through software implementation and testing, it will be possible to search for and develop possibilities for implementing such a model using neurons, synapses, and biologically based neural dynamics. This could be an easier approach to finding a neural implementation for our semantic network model than attempting to do so without a dynamic version.

If our theory for knowledge representation in the brain via semantic-network-like structures is demonstrated to be sound via one or more possible neural implementations, it will then be possible to test the theory's validity with experimental predictions.

\section{Acknowledgments}

The authors would like to thank Pamela Gibson Lake and Leslie Aderhold for helpful discussions along with Reka Albert, Jules Arginteanu and Friedrich Sommer for manuscript comments and review.

\newpage


\newpage

\begin{appendix}

\section{Mathematical viability of experience-based binder units}
\label{appendixBinderUnits}

In this appendix, we address the feasibility for the brain to used experience-based binder units for integrations it has encountered. For the purpose of analysis, we will focus the discussion on combination (and-type) integrations. We find that, while the human brain does not have the resources to utilize binder units for every \emph{possible} combination of stimulus features it \emph{can} encounter, establishing binder units for \emph{actually} encountered combinations is much more viable, especially if age-dependent unit deletion is utilized. It is difficult to assess the number of disjunctive (or-type) integrations, such as types and categories of stimuli, that an organism may learn to recognize during its lifetime, but on the assumption that this number is of an order of magnitude not too much greater than that of the number of feature combinations remembered, the brain has the resources to use experience-based binder units.

It is impossible for a brain to have pre-dedicated or innately occurring binder units for all \emph{possibly} occurring combinations of representable features.
This is particularly the case for neurally localized representations, and it is also the case in
general due to limitations on the information density per
synapse for a brain. As a trivial example, consider the
drastically reduced visual stimulus class consisting of square 5 cm  $\times$ 5 cm tiled
color patterns viewed at a distance of 30 cm---each tile being one centimeter
square and displaying one of ten unique colors.
The color patterns clearly under-sample the
space of possible visual feature combinations yet there are
$10^{25}$ of them. This is a factor of $10^{14}$ greater than the number of
neurons in the human brain, so there are not enough neurons to have a pre-existing
neurally localized binder unit for each of these integrations.

Furthermore any binder
unit for one of these color patterns, regardless of implementation,
will need to store roughly 80 bits of information (about three for each tile) to single out one pattern. In order to have a binder unit for each pattern, there will need to be $8\times10^{26}$
bits stored, which translates to 80 billion or more bits of information per human brain synapse. As 
can be seen by comparing this number to the 60 bits necessary for a synapse to pick the afferent and efferent neurons it will connect out of a pool of $10^{11}$, it is not reasonable to expect the brain to store information at this density.
Realistic stimuli involve combinations of many more
possible features. Forming a binder unit for
every combination possible cannot be done with the resources available
to the brain, and even if it could, it would be tremendously wasteful for the brain
to do things this way because almost all of the combinations
stored would never be encountered during the lifetime of an organism.

However, if binder units are created or allocated as needed to represent \emph{actually} occurring bound combinations of features, the situation is not so unmanageable. To see this, consider a human encountering stimuli at the rate of one every 300 ms, which is a typical time between saccades for a human engaging in scene perception \cite{RaynerCastelhano}. While real humans are not constantly engaged in visual scene processing---sometimes they will be engaged in auditory processing or emotional processing, etc.---the saccade rate for visual processing offers a reasonable estimate for the rate at which stimuli across modalities may be perceived as distinct and differentiable in the case that the relevant modality is being attended to closely. Of course, humans are not, with constancy, attending closely to anything at all, in which case the rate of distinct stimulus perception will likely decrease, especially when sleep is taken into account. We will therefore treat the number of 300 ms periods in a human lifetime as a strong upper bound for the number of distinct stimuli the human encounters. With a generous allowance of 95 years for a lifetime, this gives 10 billion stimuli, a number that is comparable to the total number of neurons in the human neocortex \cite{Herculano-Houzel, LarsenEtAl}.

Each of these encountered stimuli can be regarded as a single combined entity in its own right, and it might also involve on the order of one to ten differentiated sub-entities (e.g., background, foreground, table, chair, pencil, paper, etc.). We will therefore use 50 billion as the maximum number of entities encountered by a human. Given this maximum, even locally represented (giant neuron) binder units become possible provided units are pruned as they age: it becomes possible to retain binder units for a good proportion of recently occurring entities as well as a non-zero amount of entities that are as old as the system itself while keeping the total number of binder units to a small fraction of the number of neurons in the cerebral cortex.

We illustrate this with an example. It is
known that synaptic structures in the brain change on multiple time
scales, and therefore it is reasonable to suppose that binder unit
removal should happen on multiple time scales as well. A power law
probability distribution is a simple form which assumes a distribution
of binder unit deletion rates. Suppose a simple power law distribution
for binder unit preservation with elapsed time since stimulus presentation,
\begin{equation}
   p(t)=\min\left(1,(t/\tau)^{-k}\right) ,
\end{equation}
where $p(t)$ is the probability that a binder unit for a presented entity will
be present after time, $t$, has elapsed since the entity first
appears, $\tau$ is the duration of individual stimuli, which we will
treat as fixed for simplicity, and $k$ is a positive constant less
than unity. After time, $T_{\rm max}$, the number of binder units will be
\begin{equation}
n_{sim} \cdot \left( 1 + \int_\tau^{T_{\rm max}}{\frac{1}{\tau} (t/\tau)^{-k} dt}  \right)
= n_{sim} + \frac{n_{sim}}{1-k} \left((T_{\rm max}/\tau)^{1-k} - 1 \right),
\end{equation}
where $n_{sim}$ is the number of simultaneously occurring differentiated entities.

For $k=0.5$ and $n_{sim} = 5$, after 10 billion stimuli there will be about 1 million
total binder units, 200 thousand of which will correspond to entire stimuli.
By comparison, there are about 15 million granule
cells in the dentate gyrus\footnote{The dentate gyrus is part of the
  hippocampal formation and is thought to be important to the
  formation of episodic memories---which are not unlike binder units
  for encountered stimuli \cite{AlvarezSquire,Fernandez}.}  of
the human brain alone \cite{WestGundersen}. This gives a 75:1 ratio
for dentate gyrus granule cells to preserved entire stimuli (which can be thought of
as episodic memories) and a 10000:1 ratio for cerebral cortex neurons to
preserved combinations in general. The oldest binder units will have a $10^{-5}$ retention
probability, which for $\tau=300$ ms and $T_{\rm max}=$ 95 years,
corresponds to roughly 12 of the oldest stimuli preserved on average
per day's worth of input. For these parameters, most stimuli younger
than 1200 ms will be retained, and stimuli about 30 seconds old have
around 10\% chance of retention---corresponding to, on average, five for every three seconds
of input this old. The actual probability distribution for binder unit
preservation may be different than this, but this simple example
clearly shows that experience-based, neural assembly, combination binder units with
age-dependent deletion are viable for the human brain\footnote{Of course, if the
representation of binder units is not a local representation at the neural
level, even more retained bound memories are possible within the
dentate gyrus alone.}.

This discussion provides a proof-of-concept for experience-based combination binder units with age-dependent deletion in the case of humans. Minor modifications, such as a steeper age-dependent deletion curve, will be suitable to make the case for other organisms. Furthermore, the 10000:1 ratio for cerebral cortex neurons to preserved combination binder units found above leaves ample room for up to 100 times as many disjunctive binder units even if coded locally.

\newpage
\section{Selected Glossary}
\label{App:Glossary}

In this section, we provide definitions for a few important terms as they are used in this paper.

{\bf assembly (neural)}
\begin{description}
\item[ ] \emph{A set of neurons that corresponds to an idea. Assemblies can overlap.}
\end{description}

{\bf binding}
\begin{description}
\item[ ] \emph{The joining of a set of features, entities or other components into a unified whole or {\bf integration}.}
\end{description}

{\bf binder unit}
\begin{description}
\item[ ] \emph{An element of memory that serves to bind a set of features, entities or other {\bf components} into an {\bf integration}. Binder units may be implemented by \bf assemblies.}
\end{description}

{\bf component}
\begin{description}
\item[ ] \emph{An idea that contributes in some way to the meaning of another idea.}
\end{description}

{\bf concept}
\begin{description}
\item[ ] \emph{A more general idea which other, more specific ideas may implement. For example, `fox' is a concept instantiated by particular foxes. See {\bf type node}.}
\end{description}

{\bf conditional meaning}
\begin{description}
\item[ ] \emph{An idea's conditional meaning is the set of pre-conditions, including environmental events and activated ideas, which are likely to cause the idea itself to become active.}
\end{description}

{\bf consequential meaning}
\begin{description}
\item[ ] \emph{An idea's consequential meaning is the set of post-conditions, including environmental events and activated ideas, which are likely to occur due to the idea being active.}
\end{description}

{\bf criteriality}
\begin{description}
\item[ ] \emph{How important an idea or environmental event is to the meaning of another idea.}
\end{description}

{\bf entity}
\begin{description}
\item[ ] \emph{An idea corresponding to a concrete, often physical, thing such as a particular image of a fox as opposed to the concept of foxes in general.}
\end{description}

{\bf external event}
\begin{description}
\item[ ] \emph{An event in the world external to a neurally implemented semantic network which may be linked with a node in the network, thus contributing to that node's meaning.}
\end{description}

{\bf giant neuron coding}
\begin{description}
\item[ ] \emph{Coding by way of neural assemblies which are mutually exclusive.}
\end{description}

{\bf idea}
\begin{description}
\item[ ] \emph{Anything potentially recognizable by a semantic or neural network, including {\bf entities}, {\bf concepts}, {\bf components}, {\bf relationship types} and {\bf relationship instances}}.
\end{description}

{\bf instance node}
\begin{description}
\item[ ] \emph{A node in a semantic network representing an idea that is a member of a type or class of ideas recognized by the network. See {\bf type node}.}
\end{description}

{\bf integration (bound)}
\begin{description}
\item[ ] \emph{A whole involving {\bf components}, e.g., an entity that is a combination of features and sub-entities is an integration, and so is a type to which multiple entities or sub-types belong.}
\end{description}

{\bf link}
\begin{description}
\item[ ] \emph{See \bf semantic network.}
\end{description}

{\bf local coding}
\begin{description}
\item[ ] \emph{See \bf giant neuron coding.}
\end{description}

{\bf node}
\begin{description}
\item[ ] \emph{See \bf semantic network.}
\end{description}

{\bf overlapping set coding}
\begin{description}
\item[ ] \emph{Coding by way of neural assemblies which are not mutually exclusive.}
\end{description}

{\bf relational position (relationship position)}
\begin{description}
\item[ ] \emph{A characteristic role in a {\bf relationship type}. For example, `X gives Y to Z' is a relationship type that involves three relational positions: giver, given and receiver.}
\end{description}

{\bf relational position instance}
\begin{description}
\item[ ] \emph{An particular instance of a {\bf relational position} as involved in a {\bf relationship instance}, for example, the `giver' position in an instance of `giving'.} See \bf{relationship type}.
\end{description}

{\bf relationship (instance)}
\begin{description}
\item[ ] \emph{An idea that integrates two or more other ideas, expressing one aspect of the manner in which those ideas are involved with one another. See {\bf relationship type}}.
\end{description}

{\bf relationship type}
\begin{description}
\item[ ] \emph{A class to which {\bf relationship instances} belong. `X causes Y' is a relationship type. ``The moon causes tides'' is an instance of `X causes Y'. See {\bf relational position}. }
\end{description}

{\bf semantic network}
\begin{description}
\item[ ] \emph{A collection of {\bf nodes} standing for ideas, some of which are pairwise connected by {\bf links}. The structure of the network expresses meaning.}
\end{description}

{\bf type node}
\begin{description}
\item[ ] \emph{A node that integrates the meanings of multiple {\bf instance nodes} by signifying that the ideas they stand for all belong to a certain class or type. For example, spring, summer, winter and autumn are all ideas belonging to the type, `season'. Type nodes also integrate typical properties and characteristics for ideas belonging to the type, e.g., ``about three months long.''}
\end{description}

\newpage
\section{Textual Representation of Number Network}
\label{TextNumberNet}
\begin{mylisting}
\begin{verbatim}

1, 2, 3, 4, 5, 6, a, b, and n are numbers. a, b, and n are generic.

"2 succeeds 1," "3 succeeds 2," "4 succeeds 3," "5 succeeds 4," and "6 succeeds 5,"
and "succ(n) succeeds n" are all examples of succession, a relationship involving a
predecessor and a successor.

In the case of "2 succeeds 1," 2 is the successor and 1 is the predecessor.
In the case of "3 succeeds 2," 3 is the successor and 2 is the predecessor.
In the case of "4 succeeds 3," 4 is the successor and 3 is the predecessor.
In the case of "5 succeeds 4," 5 is the successor and 4 is the predecessor.
In the case of "6 succeeds 5," 6 is the successor and 5 is the predecessor.

"1+1", "2+1", "n+1", "2+2", "3+2", "4+2", "3+3", "a+b", and "b+a" are all sums.
A sum is a numerical combination involving two numbers, or terms.

For the sum, "1+1", the two terms are 1 and 1.
For "2+1", the terms are 2 and 1.
For "n+1", the terms are n and 1.
For "2+2", the terms are 2 and 2.
For "3+2", the terms are 3 and 2.
For "4+2", the terms are 4 and 2.
For "3+3", the terms are 3 and 3.
For "a+b", the terms are a and b.
For "b+a", the terms are b and a.

"1+1=2", "2+1=3", "n+1=succ(n)", "2+2=4", "3+2=5", "4+2=6", "3+3=6", and "a+b=b+a"
are all examples of equality, a relationship between two numbers or terms.

For "1+1=2", the two terms are "1+1" and 2.
For "2+1=3", the terms are "2+1" and 3.
For "n+1=succ(n)", the terms are "n+1" and succ(n).
For "2+2=4", the terms are "2+2" and 4.
For "3+2=5", the terms are "3+2" and 5.
For "4+2=6", the terms are "4+2" and 6.
For "3+3=6", the terms are "3+3" and 6.
For "a+b=b+a", the terms are "a+b" and "b+a".

\end{verbatim}
\end{mylisting}

\end{appendix}

\end{document}